\documentclass[a4paper, 11pt]{article}
\usepackage[T1]{fontenc}
\usepackage{graphicx,epsf,amssymb,amsbsy,amsfonts,amssymb,amsmath,courier,jheppub,romannum,empheq}
\def\be{\begin{equation}}
\def\ee{\end{equation}}
\def\bea{\begin{eqnarray}}
\def\eea{\end{eqnarray}}

\def\e{\epsilon}

\def\l{\lambda}
\def\l{\lambda}

\def\bg{\bar{g}}
\def\beq{\begin{eqnarray}}\def\eeq{\end{eqnarray}}
\def\ba#1\ea{\begin{align}#1\end{align}}
\def\bg#1\eg{\begin{gather}#1\end{gather}}
\def\bm#1\em{\begin{multline}#1\end{multline}}
\def\bmd#1\emd{\begin{multlined}#1\end{multlined}}
\newcommand{\ket}[1]{\left| #1 \right>} % for Dirac bras
\newcommand{\bra}[1]{\left< #1 \right|} % for Dirac kets

\def\D{\Delta}
\def\e{\epsilon}

\def\l{\lambda}

\def\({\left(}
\def\){\right)}
\def\[{\left[}
\def\]{\right]}

\def\rta{\rightarrow}

\makeatletter
\newcommand\niton{\mathrel{\m@th\mathpalette\canc@l\owns}}
\newcommand\canc@l[2]{{\ooalign{$\hfil#1/\mkern1mu\hfil$\crcr$#1#2$}}}
\makeatother

\def\l{\lambda}

\def\om{\omega}

\def\D{\Delta}

\title{BMS Symmetry of Celestial OPE}
\preprint{SAGEX-20-01-E}
\author[1,2]{Shamik~Banerjee,}
\author[3]{Sudip~Ghosh,}
\author[4]{Riccardo~Gonzo}
\affiliation[1] {Institute of Physics,\\ Sachivalaya Marg, Bhubaneshwar, India-751005}
\affiliation[2]{Homi Bhabha National Institute, Anushakti Nagar, Mumbai, India-400085}
\affiliation[3] {Okinawa Institute of Science and Technology, \\ 1919-1 Tancha, Onna-son, Okinawa 904-0495,Japan}
\affiliation[4] {School of Mathematics and Hamilton Mathematics Institute,\\ Trinity College Dublin, Ireland}
\emailAdd{banerjeeshamik.phy@gmail.com}
\emailAdd{sudip112phys@gmail.com}
\emailAdd{gonzo@maths.tcd.ie}

\abstract {In this paper we study the BMS symmetry of the celestial OPE of two positive helicity gravitons in Einstein theory in four dimensions. The celestial OPE is obtained by Mellin transforming the scattering amplitude in the (holomorphic) collinear limit. The collinear limit at leading order gives the singular term of the celestial OPE. We compute the first subleading correction to the OPE by analysing the four graviton scattering amplitude directly in Mellin space. The subleading term can be written as a linear combination of BMS descendants with the OPE coefficients determined by BMS algebra and the coefficient of the leading term in the OPE. This can be done by defining a suitable BMS primary state. We find that among the descendants, which appear at the first subleading order, there is one which is created by holomorphic supertranslation with simple \emph{pole} on the celestial sphere.}

\begin{document}
%\preprint{}
\maketitle
%\tableofcontents
\flushbottom
\section{Introduction}
Operator product expansion (OPE) plays a central role in the non-perturbative formulation of conformal field theory. OPE is the statement that when two primary operators $\phi_i$ and $\phi_j$ come close to each other (inside a correlation function) we can replace the product $\phi_i\phi_j$ by a sum over conformal families each of which contains a primary operator, say $\phi_k$ and its descendants. In general, the OPE coefficient $C_{ij}^k$ which multiplies the primary operator $\phi_k$, cannot be determined by conformal symmetry alone. But, once $C_{ij}^k$ is specified, the coefficients of all the descendants of $\phi_k$ are completely fixed by conformal invariance of the OPE. $C_{ij}^k$ is known as the structure constant of the operator algebra. The goal of the present paper is to study this aspect of the OPE in case of Celestial Conformal Field Theory. 

Celestial CFT is conjectured to be the holographic dual of quantum gravity in asymptotically flat space-time \cite{Pasterski:2016qvg,Pasterski:2017kqt,Cheung:2016iub,deBoer:2003vf,Banerjee:2018gce}. The observables of the celestial CFT are related to Mellin transformations of flat space scattering amplitudes\cite{Pasterski:2017ylz,Schreiber:2017jsr,Cardona:2017keg,Lam:2017ofc,Banerjee:2017jeg,Banerjee:2019prz,Stieberger:2018edy,Stieberger:2018onx}. Under Lorentz transformations, which act on the celestial sphere as global conformal group, Mellin amplitudes transform like correlation functions of a CFT. Now the correspondence between soft theorems \cite{Weinberg:1965nx,Cachazo:2014fwa} and Ward identities for asymptotic symmetries \cite{Strominger:2013jfa,He,Strominger:2014pwa,Barnich:2009se,Barnich:2011ct,Kapec:2014opa,Kapec:2016jld,He:2017fsb,Campiglia:2015kxa,Campiglia:2014yka,Banerjee:2018fgd,Kapec:2017gsg} show that the celestial CFT has, in fact, a much larger symmetry known as BMS\cite{Bondi:1962px,Sachs:1962zza}. The BMS group\footnote{with an abuse of notation, in this paper we call BMS group what would be appropriate to call extended BMS group} is an extension of the usual Poincar\'e group and consists of superrotations \cite{Barnich:2009se,Barnich:2011ct,Kapec:2014opa,Kapec:2016jld}, which are local conformal transformations of the celestial sphere, and supertranslations, which are local angle-dependent space-time translations at null-infinity. Due to the presence of the supertranslations, the properties of the celestial CFT are somewhat different from usual CFT. For example, BMS algebra is not a direct product of holomorphic and antiholomorphic transformations because supertranslation generators have both holomorphic and antiholomorphic weights. As a result, at least naively, we do not expect holomorphic factorisation at the level of BMS representations. This is a major difference from usual CFTs. 

A useful way to study various aspects of celestial CFT and representation theory of BMS algebra is through the construction of celestial OPE. OPE of two primary operators can be obtained by Mellin transformation of the collinear limit of flat space scattering amplitudes \cite{Pate:2019lpp,Fotopoulos:2019vac,Fan:2019emx}. In the collinear limit, at leading order an $(n+1)$ point function factorizes into an $n$ point function times a universal splitting function\cite{Bern:1998sv,Akhoury:2011kq}. By the Mellin transformation of the splitting function one obtains the leading term in the celestial OPE and the structure constant of the celestial operator algebra. It is conceivable that the subleading terms in the OPE can be generated by Mellin transforming the subleading terms in the collinear expansion\cite{Nandan:2016ohb}. Now for the celestial OPE what is remarkable is that one can obtain the structure constant by imposing a constraint coming from the subleading soft theorem in case of gluons and subsubleading soft theorem in case of gravitons\cite{Pate:2019lpp}. This suggests that owing to an unusually large amount of global symmetry, algebraic techniques \cite{Pate:2019lpp,Banerjee:2019aoy,Banerjee:2019tam,Fotopoulos:2019tpe,Law:2019glh} may play a crucial role in determining the structure of celestial correlation functions (or flat-space $S$-matrix elements). This, in particular, will require an understanding of BMS representation theory in the context of $S$-matrix theory or celestial amplitudes. 

Motivated by this, in this paper we compute the first subleading correction to the (holomorphic) collinear limit directly in the Mellin space. The subleading terms in the collinear limit give the subleading terms in the celestial OPE. We focus on the tree level four graviton scattering amplitude in Einstein theory and compute subleading OPE of two positive helicity outgoing graviton primaries. Unlike in the case of $2$-D CFT, the first correction to the leading order result contains the supertranslation descendant created by \emph{singular} supertranslation of the form $u\rightarrow u + \epsilon/z$. We also show that the subleading OPE coefficients can be derived from the BMS algebra once we define a suitable notion of BMS primary state. This suggests the possibility that just like in the case of ordinary CFT, celestial OPE also organizes itself into representations of BMS algebra with the OPE coefficients of BMS descendants determined by BMS algebra. It will be very interesting to prove or disprove this in complete generality. 
   
\section{BMS algebra}
Let us now describe BMS transformations acting on a three dimensional space with coordinates $(u,z,\bar z)$ where $u$ can be thought of as the retarded or Bondi time and $(z,\bar z)$ are the stereographic coordinates of the celestial sphere. At the end, when we derive the subleading OPE coefficients from the BMS algebra, we will restrict to the celestial sphere at $u=0$.  

BMS transformations consist of two parts, superrotation and supertranslation. Infinitesimal superrotation is the transformation given by,
\be\label{sur1}
z \rightarrow z + \epsilon z^{n+1} , \quad \bar z \rightarrow \bar z,  \quad u \rightarrow u + \frac{1}{2} \epsilon z^{n+1},  \quad n\in \mathbb Z 
\ee 
and its antiholomorphic counterpart,
\be
z \rightarrow z, \quad \bar z \rightarrow \bar z + \bar\epsilon \bar z^{n+1}, \quad u \rightarrow u + \frac{1}{2} \bar\epsilon \bar z^{n+1},  \quad n\in \mathbb Z 
\ee 
The corresponding generators are denoted by $L_n$ and $\bar L_{n}$, respectively. They satisfy the (centerless) Virasoro algebra \cite{Barnich:2009se,Barnich:2011ct}, 
\be
[L_m,L_n] = (m-n) L_{m+n}, \quad [\bar L_m,\bar L_n] = (m-n) \bar L_{m+n}
\ee
Among these, $\{L_0,L_{\pm1},\bar L_0,\bar L_{\pm1}\}$ are the generators of Lorentz or global conformal transformations.

Supertranslations act as,
\be\label{sut1}
z \rta z , \quad \bar z \rightarrow \bar z, \quad  u \rta u + \epsilon z^{a+1} \bar z^{b+1} , \quad a,b \in \mathbb Z
\ee
We denote the corresponding generators by $P_{a,b}$. They satisfy the algebra \cite{Barnich:2009se,Barnich:2011ct,Banerjee:2018fgd},
\be
[P_{a,b}, P_{a',b'}] =0 
\ee
In this case, $\{P_{-1,-1},P_{-1,0},P_{0,-1},P_{0,0}\}$ are the generators of global space-time translations.  

The commutator algebra between supertranslation and superrotation is given by \cite{Barnich:2009se,Barnich:2011ct},
\be\label{trc}
[L_n, P_{a,b}] = \bigg(\frac{n-1}{2} - a\bigg) P_{a+n,b} \qquad   [\bar L_n, P_{a,b}] = \bigg(\frac{n-1}{2} - b\bigg) P_{a,b+n}
\ee

\subsection{The Transformation of Fields}
Under an infinitesimal conformal transformation \eqref{sur1}, the primary field $\phi_{h,\bar h}(u,z,\bar z)$ of weight $(h,\bar h)$ transforms as \cite{Banerjee:2018gce},

\be\label{sur}
\delta \phi_{h,\bar h}(u,z,\bar z) = \epsilon \left[ z^{n+1} \partial + \left(n+1\right) \left(h+ \frac{1}{2} u \partial_u\right) z^n \right] \phi_{h,\bar h}(u,z,\bar z)
\ee
Similarly for antiholomorphic transformation \cite{Banerjee:2018gce},

\be\label{sur2}
\delta \phi_{h,\bar h}(u,z,\bar z) = \epsilon \bigg[ \bar z^{n+1} \bar\partial + (n+1) \left(\bar h+ \frac{1}{2} u \partial_u\right) \bar z^n \bigg] \phi_{h,\bar h}(u,z,\bar z)
\ee

For infinitesimal supertranslation given by \eqref{sut1} the transformation of the primary field is given by \cite{Banerjee:2018gce},
\be\label{sut}
\delta \phi_{h,\bar h}(u,z,\bar z) = \epsilon z^{a+1} \bar z^{b+1}  \partial_u \phi_{h,\bar h}(u,z,\bar z)
\ee

At this point we would like to mention one useful point. From the transformation laws \eqref{sur}, \eqref{sur2} and \eqref{sut} it is easy to check that if $\phi_{h,\bar h}(u,z,\bar z)$ is a primary then so is $(\partial/\partial u)^n \phi_{h,\bar h}(u,z,\bar z)$, with weight $(h+ n/2, \bar h + n/2)$. 

\section{Superrotation and Supertranslation Ward Identities} 
In celestial CFT correlation functions of the two dimensional primary operator $\phi_{h,\bar h}(z,\bar z)$ are defined as Mellin transformation of flat space $S$-matrix elements \cite{Pasterski:2016qvg,Pasterski:2017kqt},
\be\label{mellin}
\langle \prod_{i=1}^n\phi_{h_i,\bar h_i}(z_i,\bar z_i)\rangle = \prod_{i=1}^{n} \int_{0}^{\infty} d\om_i \ \om_i^{i\l_i} \mathcal{A}(\{p_{i}(\om_i,z_i,\bar z_i),\sigma_i\})
\ee  
where $\epsilon_i=\pm 1$ for outgoing and incoming particles, respectively. The null momentum $p(\om,z,\bar z)$ is parametrised as,
\be
p(\om,z,\bar z) = \om (1+z\bar z, z+\bar z, -i(z-\bar z), 1-z\bar z)
\ee
and $\sigma$ denotes the helicity of the particle. Under (Lorentz) global conformal transformation, the L.H.S of \eqref{mellin} transforms as the correlation function of primary operators of weight $(h_i,\bar h_i)$, given by 
\be
h_i = \frac{1+i\l_i + \sigma_i}{2}, \qquad \bar h_i = \frac{1+i\l_i -\sigma_i}{2}
\ee
The action of global space-time translation on \eqref{mellin} was studied in \cite{Stieberger:2018onx}.

The two dimensional field $\phi_{h,\bar h}(z,\bar z)$ is the restriction of the three dimensional field $\phi_{h,\bar h}(u,z,\bar z)$ to the $u=0$ celestial sphere. Correlation function of the three dimensional fields $\phi_{h,\bar h}(u,z,\bar z)$ is defined as \cite{Banerjee:2018gce}, 
\be\label{modmellin}
\langle{\prod_{i=1}^n\phi_{h_i,\bar h_i}(u_i,z_i,\bar z_i)}\rangle \ = \prod_{i=1}^{n} \int_{0}^{\infty} d\om_i \ \om_i^{i\l_i} e^{-i\epsilon_i\om_i u_i} \mathcal{A}(\{p_{i}(\om_i,z_i,\bar z_i),\sigma_i\})
\ee
Under (Lorentz) global conformal transformation, the L.H.S of \eqref{modmellin} transforms as,
\be
\langle\prod_{i=1}^n\phi_{h_i,\bar h_i}(u_i,z_i,\bar z_i)\rangle \  = \prod_{j=1}^{n} \frac{1}{(cz_j + d)^{2h_j}} \frac{1}{(\bar c \bar z_j + \bar d)^{2\bar h_j}} \langle\prod_{i=1}^n \phi_{h_i,\bar h_i}\bigg(\frac{u_i}{|cz_i + d|^2} \ , \frac{az_i+b}{cz_i+d} \ ,\frac{\bar a \bar z_i + \bar b}{\bar c \bar z_i + \bar d} \bigg)\rangle
\ee 
Similarly, if we do a global space-time translation under which $u\rightarrow u + a + bz +\bar b \bar z + cz \bar z$, with $(z,\bar z)$ remaining fixed, the correlation function \eqref{modmellin} is invariant. Let us now discuss the transformation law of the correlation functions under local BMS transformations which are captured by BMS Ward identities. 

It is well known that Cachazo-Strominger subleading soft graviton theorem \cite{Cachazo:2014fwa} is equivalent to the (superrotation) conformal Ward identity \cite{Kapec:2016jld},
\be\label{cs}
\langle T(z) \prod_{i=1}^n \phi_{h_i, \bar h_i}(u_i,z_i,\bar z_i)\rangle = \sum_{k=1}^n \bigg( \frac{h_k + \frac{1}{2} u_k \partial_{u_k}}{(z-z_k)^2} + \frac{1}{z-z_k}\frac{\partial}{\partial z_k} \bigg) \langle\prod_{i=1}^n \phi_{h_i, \bar h_i}(u_i,z_i,\bar z_i)\rangle
\ee
where the stress tensor $T(z)$ can be constructed as the shadow of the subleading soft graviton. In \cite{Kapec:2016jld} the Ward-identity was derived for the two dimensional fields $\phi_{h,\bar h}(z,\bar z)$, but the same derivation can be easily repeated for the fields $\phi_{h,\bar h}(u,z,\bar z)$ and one obtains \eqref{cs}. For details please see the appendix. It is important to note that the stress tensor $T(z)$ does not depend on the time coordinate $u$ because it is constructed from a soft graviton and in the soft limit the time coordinate decouples. 

The singular terms in the OPE between the stress tensor $T(z)$ and the primary $\phi_{h,\bar h}(u,w,\bar w)$ are given by \cite{Fotopoulos:2019tpe},
\be\label{opetp}
T(z) \phi_{h,\bar h}(u,w,\bar w)  \sim  \frac{h + \frac{1}{2} u \partial_{u}}{(z-w)^2} \phi_{h,\bar h}(u,w,\bar w) + \frac{1}{z-w} \frac{\partial}{\partial w} \phi_{h,\bar h}(u,w,\bar w)
\ee
This is consistent with the transformation law \eqref{sur} which one can check by using the standard $2$-D CFT method. The OPE \eqref{opetp} gives the commutation relation, 
\be\label{cv}
[L_n , \phi_{h,\bar h}(u,z,\bar z)] = \bigg[ z^{n+1} \partial + (n+1) \left(h+ \frac{1}{2} u \partial_u\right) z^n \bigg] \phi_{h,\bar h}(u,z,\bar z)
\ee
where the Virasoro generator $L_n$ is defined in the usual manner as,
\be
L_n = \oint_{c_0} dz z^{n+1} T(z)
\ee
with $c_0$ defined as a contour around $z=0$. 

Similarly, Weinberg's soft graviton theorem \cite{Weinberg:1965nx} is equivalent to the supertranslation Ward-identity given by \cite{Strominger:2013jfa,He,Banerjee:2018fgd},
\be\label{supward1}
\langle P(z)\prod_{i=1}^n \phi_{h_i, \bar h_i}(u_i,z_i,\bar z_i)\rangle = \sum_{k=1}^n \frac{1}{z-z_k} i\frac{\partial}{\partial u_k} \langle\prod_{i=1}^n \phi_{h_i, \bar h_i}(u_i,z_i,\bar z_i)\rangle
\ee
Here $P(z)$ is the supertranslation current and can be written as, $P(z) = - \lim_{i\l \rightarrow 0}i\l\bar\partial G^+_{\D=1+i\l}$, where $G^+_{\D}$ is the positive helicity graviton primary of weight $\D (=h+\bar h)$. Again, $P(z)$ has no $u$ dependence. 

The singular term in the OPE between $P(z)$ and a matter primary $\phi_{h,\bar h}(u,z,\bar z)$ is given by,
\be
P(z) \phi_{h,\bar h}(u,w,\bar w) \sim \frac{1}{z-w} i \frac{\partial}{\partial u} \phi_{h,\bar h}(u,w,\bar w)
\ee
This OPE is equivalent to the commutation relation,
\be\label{csm}
[P_{a,-1},\phi_{h,\bar h}(u,z,\bar z)] = z^{a+1}  i \partial_u \phi_{h,\bar h}(u,z,\bar z)
\ee
where the supertranslation generator $P_{a,-1}$, defined as \cite{Strominger:2013jfa},
\be
P_{a,-1} = \oint_{c_0} dz z^{a+1} P(z)
\ee
generates the holomorphic supertranslation $u \rightarrow u + \epsilon z^a$. The commutation relation \eqref{csm} shows that, 
\be
[P_{a,-1},\phi_{h,\bar h}(0)] = 0 , \qquad    a > -1
\ee
From the BMS commutation relation \eqref{trc} one can check that the supertranslation generator $P_{a,-1}$ has weight $(-a-\frac{1}{2},\frac{1}{2})$. So for $a>-1$ the holomorphic weight of $P_{a,-1}$ is negative and it annihilates the primary operator $\phi_{h,\bar h}(0)$. 

%An identical discussion goes through for antiholomorphic generators. 

\section{BMS Descendants and Their Correlators}
\subsection{Supertranslation}
Let us first consider the (holomorphic) supertranslation descendants. 

Let us assume that the standard CFT form for the OPE between the supertranslation current $P(z)$ and a matter primary $\phi_{h,\bar h}(u,z,\bar z)$ holds, i.e, 
\be\label{opesm}
\begin{gathered}
P(w) \phi_{h,\bar h}(u,z,\bar z) = \frac{(P_{-1,-1}\phi_{h,\bar h})(u,z,\bar z)}{w-z} + (P_{-2,-1}\phi_{h,\bar h})(u,z,\bar z) \\+ (w-z) (P_{-3,-1}\phi_{h,\bar h})(u,z,\bar z) + (w-z)^2 (P_{-4,-1}\phi_{h,\bar h})(u,z,\bar z) + ......
\end{gathered} 
\ee
where the leading term is given by,
\be
(P_{-1,-1}\phi_{h,\bar h})(u,z,\bar z) = i\frac{\partial}{\partial u} \phi_{h,\bar h}(u,z,\bar z)
\ee
From the conformal transformation laws \eqref{sur} and \eqref{sur2} of primary fields, we know that $i\frac{\partial}{\partial u} \phi_{h,\bar h}(u,z,\bar z)$ also transforms like a primary field of weight $(h+1/2,\bar h+1/2)$. We denote this field by
\be\label{pd}
\phi_{h+1/2,\bar h+1/2}(u,z,\bar z) = i\frac{\partial}{\partial u} \phi_{h,\bar h}(u,z,\bar z) 
\ee

The nonsingular terms of the OPE \eqref{opesm} define the (holomorphic) supertranslation descendants $\bigg\{(P_{-n,-1}\phi_{h,\bar h})(u_0,z_0,\bar z_0)\bigg\}_{n\ge 2}$ which are new local fields created by singular supertranslations of the form,
\be
u \rightarrow u + \frac{\e}{(z-z_0)^{n-1}}
\ee
The descendants can be defined by the usual contour integral formula,
\be
(P_{-a,-1}\phi_{h,\bar h})(u,z,\bar z) = \oint_{c_z} dw \frac{1}{(w-z)^{a-1}} P(w) \phi_{h,\bar h}(u,z,\bar z)
\ee
which follows from \eqref{opesm}. Here $c_z$ is a contour around $w=z$. Later in the paper we will explicitly verify the existence of these descendants by taking the leading conformal soft limit of tree-level four graviton scattering amplitude in Einstein theory. 

We now need to find out correlation functions with the insertion of the descendants $P_{-a,-1}\phi_{h,\bar h}$, i.e, correlators of the form $\langle(P_{-a,-1}\phi_{h,\bar h})(u,z,\bar z)\prod_{i=1}^{n}\phi_{h_i,\bar h_i}(u_i,z_i,\bar z_i)\rangle$. This can be computed in the standard way by using the Ward identity \eqref{supward1},
\begin{align}
\langle P(w)\phi_{h,\bar h}(u,z,\bar z)\prod_{i=1}^{n}\phi_{h_i,\bar h_i}(u_i,z_i,\bar z_i)\rangle  
=  &\sum_{k=1}^n \frac{1}{w-z_k} i\frac{\partial}{\partial u_k}\langle\phi_{h,\bar h}(u,z,\bar z)\prod_{i=1}^{n}\phi_{h_i,\bar h_i}(u_i,z_i,\bar z_i)\rangle \nonumber\\ &+ \frac{1}{w-z} i\frac{\partial}{\partial u}\langle\phi_{h,\bar h}(u,z,\bar z)\prod_{i=1}^{n}\phi_{h_i,\bar h_i}(u_i,z_i,\bar z_i)\rangle
\end{align}
and taking the limit $w \rightarrow z$. %The position of the supertranslation current $P(w)$ can be varied independently inside the correlation function because, being a soft operator, $P(w)$ does not contribute to the energy-momentum conservation.
In this limit we can use the OPE \eqref{opesm} and obtain,
\be\label{superd}
\begin{gathered}
\hspace{-20pt}\boxed{\langle(P_{-a,-1}\phi_{h,\bar h})(u,z,\bar z)\prod_{i=1}^{n}\phi_{h_i,\bar h_i}(u_i,z_i,\bar z_i)\rangle = - \sum_{k=1}^n \frac{1}{(z_k - z)^{a-1}} i \frac{\partial}{\partial u_k} \langle\phi_{h,\bar h}(u,z,\bar z)\prod_{i=1}^{n}\phi_{h_i,\bar h_i}(u_i,z_i,\bar z_i)\rangle} \\ = \mathcal P_{-a,-1}(z) \langle\phi_{h,\bar h}(u,z,\bar z)\prod_{i=1}^{n}\phi_{h_i,\bar h_i}(u_i,z_i,\bar z_i)\rangle
\end{gathered}
\ee
where the differential operator $\mathcal P_{-a,-1}(z)$ acting on correlation functions of primary operators is defined as,
\be
\mathcal P_{-a,-1}(z) =  - \sum_{k=1}^n \frac{1}{(z_k - z)^{a-1}} i \frac{\partial}{\partial u_k}
\ee
The rest of the (holomorphic) supertranslation descendants are of the form \\ $P_{-i_1,-1}P_{-i_2,-1}....P_{-i_n,-1}\phi_{h,\bar h}(u,z,\bar z)$ where $i_1\ge i_2\ge...\ge i_n\ge 1$. The mixed supertranslation descendants of the form $P_{-a,-b}\phi_{h,\bar h}(u,z,\bar z)$, where $a,b > 1$, will not appear in the OPE to the first subleading order and so we leave their discussion to future work. 

Mellin transform of graviton scattering amplitudes in string theory \cite{Stieberger:2018edy} is well defined without the time coordinate $u$. In this case the correlation function with the insertion of a holomorphic supertranslation descendant is given by a simple change of \eqref{superd},
\be\label{nou}
\begin{gathered}
\langle(P_{-a,-1}\phi_{h,\bar h})(z,\bar z;\epsilon)\prod_{i=1}^{n}\phi_{h_i,\bar h_i}(z_i,\bar z_i;\epsilon_i)\rangle = - \sum_{k=1}^n \frac{\epsilon_k \widetilde P_k }{(z_k - z)^{a-1}} \langle\phi_{h,\bar h}(z,\bar z;\epsilon)\prod_{i=1}^{n} \phi_{h_i,\bar h_i}(z_i,\bar z_i;\epsilon_i)\rangle \\ = \mathcal P_{-a,-1}(z) \langle \phi_{h,\bar h}(z,\bar z;\epsilon)\prod_{i=1}^{n}\phi_{h_i,\bar h_i}(z_i,\bar z_i;\epsilon_i)\rangle \\ 
\end{gathered}
\ee
where $\epsilon=\pm 1$ for an outgoing and incoming particle, respectively and 
\be
\widetilde P_i \ \phi_{h_j,\bar h_j}(z_j,\bar z_j;\epsilon_j) = \phi_{h_j+1/2,\bar h_j+1/2}(z_j,\bar z_j;\epsilon_j) \ \delta_{ij}
\ee 

\subsection{Superrotation or Virasoro}

The discussion of superrotation or Virasoro descendants is identical to that in $2$-D CFT. The correlation function with the insertion of $(L_{-n}\phi_{h,\bar h})(u,z,\bar z)$ is given by,
\be
\langle(L_{-n}\phi_{h,\bar h}(u,z,\bar z))\prod_{i=1}^p\phi_{h_i,\bar h_i}(u_i,z_i,\bar z_i)\rangle = \mathcal L_{-n}(z) \langle\phi_{h,\bar h}(u,z,\bar z)\prod_{i=1}^p\phi_{h_i,\bar h_i}(u_i,z_i,\bar z_i)\rangle
\ee
where
\be
\mathcal L_{-n}(z) = - \sum_{i=1}^{p} \bigg( (1-n)\frac{h_i + \frac{1}{2}u_i\frac{\partial}{\partial u_i}}{(z_i - z)^n}  + \frac{1}{(z_i -z)^{n-1}}\frac{\partial}{\partial z_i}\bigg)
\ee
This can be obtained by assuming the following OPE between the stress tensor $T(z)$ and the primary field $\phi_{h,\bar h}(u,z,\bar z)$,
\be
\begin{gathered}
T(z) \phi_{h,\bar h}(u,w,\bar w) = \frac{(h+ \frac{1}{2} u\partial_u)\phi_{h,\bar h}(u,w,\bar w)}{(z-w)^2} + \frac{\partial\phi_{h,\bar h}(u,w,\bar w)}{z-w} + (L_{-2}\phi_{h,\bar h})(u,w,\bar w) \\ + (z-w) (L_{-3}\phi_{h,\bar h})(u,w,\bar w) + ......
\end{gathered}
\ee
In the absence of the time coordinate $u$, superrotation transformations act on the primaries $\phi_{h,\bar h}(z,\bar z)$ exactly in the same way as local conformal transformations act on Virasoro primaries in $2$-D CFT. 

\section{OPE From Four Graviton Scattering Amplitude in Einstein Theory}
In this section and the following we denote a graviton primary operator of scaling dimension $\D(=1+i\l)$ by $G_{\D}^{\pm}$ where $\pm$ is the helicity. The simplified notation $G_{\D_i}^{\pm}(i)$ means that the primary operator is inserted at the point $(u_i,z_i,\bar z_i)$. \\ 

For simplicity we focus on the four graviton tree-level scattering amplitude in Einstein theory, given by \footnote{An $n$-graviton amplitude is multiplied by a factor of $(\frac{\kappa}{2})^{n-2}$ where $\kappa = \sqrt{32\pi G_{N}}$. To simplify the formulas we work in units where $\kappa = 2$.}
\be
\begin{gathered}
M_4(1^-2^-3^+4^+) \\ = \frac{\langle12\rangle^7[12]}{\langle13\rangle \langle14\rangle\langle23\rangle\langle24\rangle\langle34\rangle^2} \ \delta^4 (\omega_1q(z_1,\bar z_1)+\omega_2q(z_2,\bar z_2)-\omega_3q(z_3,\bar z_3)-\omega_4q(z_4,\bar z_4)) \\
= - 4 \frac{\om_1^2\om_2^2}{\om_3\om_4}\frac{z_{12}^6 \bar z_{34}}{z_{13}z_{14}z_{23}z_{24}z_{34}} \ \delta^4 (\omega_1q(z_1,\bar z_1)+\omega_2q(z_2,\bar z_2)-\omega_3q(z_3,\bar z_3)-\omega_4q(z_4,\bar z_4))
\end{gathered}
\ee
where $(1,2)$ are incoming and $(3,4)$ are outgoing. We have also used the relations\footnote{We work in split signature and parametrize a null momentum $p$ as $p= \omega q(z,\bar z)= \omega(1+z\bar z, z+\bar z, z-\bar z,1-z\bar z)$.}
\be
\langle ij\rangle = -2 \epsilon_i \epsilon_j \sqrt{\om_i\om_j} \ z_{ij}, \qquad  [ij] = 2 \sqrt{\om_i\om_j} \ \bar z_{ij}
\ee
where $\epsilon_i = \pm1$ for an outgoing and an incoming particle, respectively. As in \cite{Pate:2019lpp}, we work in split signature so that we can treat $z$ and $\bar z$ as independent real variables. It is also important that in split signature there is a non-zero three point function which is crucial for our purpose.  

In order to facilitate the (holomorphic) OPE expansion as $z_3 \rightarrow z_4$, with $\bar z_{34}$ held fixed, we write the momentum-conserving delta function as,
\be
\begin{gathered}
\delta^4 (\omega_1q(z_1,\bar z_1)+\omega_2q(z_2,\bar z_2)-\omega_3q(z_3,\bar z_3)-\omega_4q(z_4,\bar z_4)) \\ = \frac{1}{4\om_1^*\om_2^*} \frac{1}{z_{12}^2} \delta (\om_1 - \om_1^*) \delta(\om_2 - \om_2^*) \ \delta\left(\bar z_{14} + \frac{\om_3}{\om_1^*} \frac{z_{23}}{z_{12}} \bar z_{34}\right) \delta \left(\bar z_{24} - \frac{\om_3}{\om_2^*} \frac{z_{13}}{z_{12}} \bar z_{34}\right) 
\end{gathered}
\ee
where
\be
\om_1^* = - \frac{z_{24}}{z_{12}} (\om_3 + \om_4) + \frac{z_{34}}{z_{12}}\om_3
\ee
\be
\om_2^* =  \frac{z_{14}}{z_{12}} (\om_3 + \om_4) - \frac{z_{34}}{z_{12}}\om_3
\ee
Now we make a change of variable 
\be
\om = \om_3 + \om_4, \quad \om_3 = t\om , \quad \om_4 = (1-t)\om, \qquad 0\le t\le 1
\ee
In terms of these new variables the Mellin amplitude can be computed easily and is given by, 
\be\label{OPEE}
\begin{gathered}
\widetilde M_4(1^-2^-3^+4^+) =  \langle G^-_{\D_1}(1)G^-_{\D_2}(2)G^{+}_{\D_3}(3)G^+_{\D_4}(4)\rangle  \\ = - \frac{\bar z_{34}}{z_{34}} \ \Bigg[ \frac{z_{12}^2}{z_{23}z_{31}} \Theta\bigg(\frac{z_{42}}{z_{12}}\bigg) \Theta\bigg(\frac{z_{14}}{z_{12}}\bigg) \bigg(\frac{z_{42}}{z_{12}}\bigg)^{i\lambda_1} \bigg(\frac{z_{14}}{z_{12}}\bigg)^{i\lambda_2} 
 \frac{\Gamma(2+i\sum_{i=1}^4\lambda)}{\bigg\{i\bigg(u_1\frac{z_{24}}{z_{12}} + u_2 \frac{z_{41}}{z_{12}} + u_4 \bigg)\bigg\}^{2+ i\sum_{i=1}^4\lambda_i}} \\
\times \int_{0}^{1} dt \ t^{i\lambda_3 -1}  \ (1-t)^{i\lambda_4 -1} \  \  \ \bigg( 1 - \frac{u_{12}z_{34} - u_{34}z_{12}}{u_1 z_{24} + u_4 z_{12} + u_2 z_{41}} t \bigg)^{-2 - i\sum_{i=1}^4\lambda_i} \\
 \times \bigg(1 - \frac{z_{34}}{z_{24}}t\bigg)^{1+i\lambda_1} \delta\bigg(\bar z_{14} - t \bigg(1- \frac{z_{34}}{z_{24}} t\bigg)^{-1} \bar z_{34} + t \bigg(1- \frac{z_{34}}{z_{24}} t\bigg)^{-1} \frac{1}{z_{24}} z_{34} \bar z_{34} \bigg) \\
\times \bigg(1 - \frac{z_{34}}{z_{14}}t\bigg)^{1+i\lambda_2} \delta\bigg(\bar z_{24} - t \bigg(1- \frac{z_{34}}{z_{14}} t\bigg)^{-1} \bar z_{34} + t \bigg(1- \frac{z_{34}}{z_{14}} t\bigg)^{-1} \frac{1}{z_{14}} z_{34} \bar z_{34} \bigg) \Bigg]
\end{gathered}
\ee 
In writing down this integral we have assumed the OPE limit, i.e, $|z_{34}| \ll |z_{13}|,|z_{23}|,|z_{14}|,|z_{24}|$. The details of the derivation are given in the Appendix.

\subsection{Leading Term of the OPE} 
We can see from \eqref{OPEE} that the expression inside the bracket multiplying the term $\frac{\bar z_{34}}{z_{34}}$ is finite in the limit $z_{34},\bar z_{34},u_{34}\rightarrow 0$ and so we can Taylor expand around $z_{34}=\bar z_{34}=u_{34} = 0$. The leading term in the expansion is given by,
\be\label{lead}
\begin{gathered}
\widetilde M_4(1^-2^-3^+4^+)  \\ = - \frac{\bar z_{34}}{z_{34}} \Bigg[ \frac{z_{12}^2}{z_{24}z_{41}} \Theta\bigg(\frac{z_{42}}{z_{12}}\bigg) \Theta\bigg(\frac{z_{14}}{z_{12}}\bigg) \bigg(\frac{z_{42}}{z_{12}}\bigg)^{i\lambda_1} \bigg(\frac{z_{14}}{z_{12}}\bigg)^{i\lambda_2} 
 \frac{\Gamma(2+i\sum_{i=1}^4\lambda_i)}{\bigg\{i\bigg(u_1\frac{z_{24}}{z_{12}} + u_2 \frac{z_{41}}{z_{12}} + u_4 \bigg)\bigg\}^{2+ i\sum_{i=1}^4\lambda_i}} \\
\times \delta(\bar z_{14})\delta(\bar z_{24}) \int_{0}^{1} dt \ t^{i\lambda_3 -1}  \ (1-t)^{i\lambda_4 -1} \Bigg] + ......
\end{gathered}
\ee 
Now the three graviton amplitude in Mellin space, when there are two negative helicity incoming gravitons at $1$ and $2$ with weights $\D_1(=1+i\l_1)$ and $\D_2(=1+i\l_2)$ and one positive helicity outgoing graviton at $4$ with weight $\D_3 + \D_4 -1(=1+i\l_3+i\l_4)$, is given by,
\be\label{3-point}
\begin{gathered}
\widetilde M_3 (1^- 2^- 4^+) = \langle G^-_{\D_1}(1)G^-_{\D_2}(2)G^+_{\D_3 + \D_4 -1}(4)\rangle \\ = \Theta\bigg(\frac{z_{42}}{z_{12}}\bigg) \Theta\bigg(\frac{z_{14}}{{z_{12}}}\bigg) \delta(\bar z_{14})\delta(\bar z_{24}) \frac{z_{12}^2}{z_{24}{z_{41}}} \bigg(\frac{z_{42}}{z_{12}}\bigg)^{i\l_1} \bigg(\frac{z_{14}}{z_{12}}\bigg)^{i\l_2}  \frac{\Gamma(1+ i\sum_{i=1}^4\l_i)}{\bigg\{ i \bigg(u_1 \frac{z_{24}}{z_{12}} + u_2 \frac{z_{41}}{z_{12}} + u_4 \bigg)\bigg\}^{1+i\sum_{i=1}^4\l_i}}
\end{gathered}
\ee
Using \eqref{3-point} we can write the leading term \eqref{lead} in the four point function as,
\be
\widetilde M_4(1^-2^-3^+4^+) = - B(i\l_3,i\l_4) \frac{\bar z_{34}}{z_{34}} \ i\frac{\partial}{\partial u_4} \widetilde M_3 (1^- 2^- 4^+) + ......
\ee 
where $B(p,q)$ is the Euler beta function. This leading term corresponds to the leading term in the OPE given by,
\be\label{L}
\begin{gathered}
G^+_{\D_3} (3) G^{+}_{\D_4}(4) = - B(\D_3 -1, \D_4 -1) \frac{\bar z_{34}}{z_{34}} \ i\frac{\partial}{\partial u_4} G^+_{\D_3 + \D_4 -1}(4) + ...... \\ = - B(\D_3 -1, \D_4 -1) \frac{\bar z_{34}}{z_{34}} \ G^+_{\D_3 + \D_4}(4) + ......
\end{gathered}
\ee
where $\D_i = 1+ i \l_i$. In writing the last line of \eqref{L} we have used the fact that $i\frac{\partial}{\partial u_4} G^+_{\D_3 + \D_4 -1}(4)$ is a primary with weight $(\D_3 + \D_4)$. The leading answer \eqref{L} matches with \cite{Pate:2019lpp}. 

\section{Leading Conformal Soft Limit and Holomorphic Supertranslation Descendants}
In the leading conformal soft limit $i\l_3\rightarrow 0$ \cite{Pate:2019mfs,Nandan:2019jas,Donnay:2018neh,Adamo:2019ipt,Puhm:2019zbl,Guevara:2019ypd,Banerjee:2019prz} the amplitude \eqref{OPEE} simplifies and is given by,

\be
\begin{gathered}
\lim_{i\l_3\rightarrow 0} i\l_3 \widetilde M_4(1^-2^-3^+4^+) \\ = - \ \frac{\bar z_{34}}{z_{34}} \frac{z_{12}^2}{z_{23}z_{31}} \Theta\bigg(\frac{z_{42}}{z_{12}}\bigg) \Theta\bigg(\frac{z_{14}}{z_{12}}\bigg) \bigg(\frac{z_{42}}{z_{12}}\bigg)^{i\lambda_1} \bigg(\frac{z_{14}}{z_{12}}\bigg)^{i\lambda_2} \delta(\bar z_{14}) \delta(\bar z_{24}) \\ \times \frac{\Gamma(2+i(\l_1+\l_2+\l_4))}{\bigg\{i\bigg(u_1\frac{z_{24}}{z_{12}} + u_2 \frac{z_{41}}{z_{12}} + u_4 \bigg)\bigg\}^{2+ i(\l_1+\l_2+\l_4)}} 
\end{gathered}
\ee 
This can be rewritten as,
\be\label{cs1}
\lim_{i\l_3\rightarrow 0} i\l_3 \widetilde M_4(1^-2^-3^+4^+) = - \frac{\bar z_{34}}{z_{34}} \bigg(1- \frac{z_{34}}{z_{24}}\bigg)^{-1} \bigg( 1 - \frac{z_{34}}{z_{14}}\bigg)^{-1} i \frac{\partial}{\partial u_4} \widetilde{\mathcal M}_3(1^-2^-4^+)
\ee
where the three point function $\widetilde{\mathcal M}_3(1^-2^-4^+)$ is now given by,
\be
\widetilde{\mathcal M}_3(1^-2^-4^+) = \langle G^-_{\D_1}(1)G^-_{\D_2}(2)G^+_{\D_4}(4)\rangle
\ee
Now we expand the R.H.S of \eqref{cs1} in powers of $z_{34}$, 
\be
\begin{gathered}
\bigg(1- \frac{z_{34}}{z_{24}}\bigg)^{-1} \bigg( 1 - \frac{z_{34}}{z_{14}}\bigg)^{-1} \\ = \sum_{n=0}^\infty z_{34}^n \bigg( \frac{1}{z_{24}^n} + \frac{1}{z_{24}^{n-1}z_{14}} + ...... + \frac{1}{z_{24}z_{14}^{n-1}} + \frac{1}{z_{14}^n} \bigg) = \sum_{n=0}^\infty \frac{z_{34}^n}{z_{12}} \frac{z_{14}^{n+1} - z_{24}^{n+1}}{z_{14}^n z_{24}^n} \\ = \sum_{n=0}^\infty z_{34}^n \bigg( \frac{1}{z_{24}^n} \frac{z_{14}}{z_{12}}  - \frac{1}{z_{14}^n} \frac{z_{24}}{z_{12}}\bigg)
\end{gathered}
\ee
So, 
\be
\lim_{i\l_3\rightarrow 0} i\l_3\widetilde M_4(1^-2^-3^+4^+) = - \frac{\bar z_{34}}{z_{34}} \sum_{n=0}^\infty z_{34}^n \bigg( \frac{1}{z_{24}^n} \frac{z_{14}}{z_{12}}  - \frac{1}{z_{14}^n} \frac{z_{24}}{z_{12}}\bigg) \ i \frac{\partial}{\partial u_4} \widetilde{\mathcal M}_3(1^-2^-4^+)
\ee
Now we have the following relations,
\be
i \frac{\partial}{\partial u_1} \widetilde{\mathcal M}_3(1^-2^-4^+) = \frac{z_{24}}{z_{12}} i \frac{\partial}{\partial u_4} \widetilde{\mathcal M}_3(1^-2^-4^+)
\ee
\be
i \frac{\partial}{\partial u_2} \widetilde{\mathcal M}_3(1^-2^-4^+) = - \frac{z_{14}}{z_{12}} i \frac{\partial}{\partial u_4} \widetilde{\mathcal M}_3(1^-2^-4^+)
\ee
Using them we can write,
\be
\begin{gathered}
\lim_{i\l_3\rightarrow 0} i\l_3\widetilde M_4(1^-2^-3^+4^+) = \lim_{i\l_3\rightarrow 0}\langle G^-_{\D_1}(1)G^-_{\D_2}(2)G^{+}_{\D_3}(3)G^+_{\D_4}(4)\rangle \\ = - \bar z_{34} \sum_{n=0}^{\infty} z_{34}^{n-1} \bigg\{ -\bigg( \frac{1}{z_{24}^n} i \frac{\partial}{\partial u_2} +  \frac{1}{z_{14}^n} i \frac{\partial}{\partial u_1}\bigg)\bigg\} \widetilde{\mathcal M}_3(1^-2^-4^+) \\ = - \bar z_{34} \sum_{n=0}^{\infty} z_{34}^{n-1} \bigg\{ -\bigg( \frac{1}{z_{24}^n} i \frac{\partial}{\partial u_2} +  \frac{1}{z_{14}^n} i \frac{\partial}{\partial u_1}\bigg)\bigg\} \langle G^-_{\D_1}(1)G^-_{\D_2}(2)G^+_{\D_4}(4)\rangle 
\end{gathered}
\ee
This can be written in a more suggestive form as, 
\be\label{ssuper}
\begin{gathered}
\lim_{i\l_3\rightarrow 0} \langle G^-_{\D_1}(1)G^-_{\D_2}(2)\big(-i\l_3\bar\partial_3G^{+}_{\D_3}(3)\big)G^+_{\D_4}(4)\rangle \ = \ \langle G^-_{\D_1}(1)G^-_{\D_2}(2) P(z_3) G^+_{\D_4}(4)\rangle \\ = \sum_{n=0}^{\infty} z_{34}^{n-1} \bigg\{ -\bigg( \frac{1}{z_{24}^n} i \frac{\partial}{\partial u_2} +  \frac{1}{z_{14}^n} i \frac{\partial}{\partial u_1}\bigg)\bigg\} \langle G^-_{\D_1}(1)G^-_{\D_2}(2)G^+_{\D_4}(4)\rangle \\ 
= \sum_{n=0}^{\infty} z_{34}^{n-1}\mathcal P_{-n-1,-1} \langle G^-_{\D_1}(1)G^-_{\D_2}(2)G^+_{\D_4}(4)\rangle \\
= \sum_{n=0}^{\infty} z_{34}^{n-1} \langle G^-_{\D_1}(1)G^-_{\D_2}(2)(P_{-n-1,-1}G^+_{\D_4})(4)\rangle
\end{gathered}
\ee
where we have used \eqref{superd} and the definition of the supertranslation current \footnote{Since we are working in units with $\kappa=\sqrt{32\pi G_N}=2$, the multiplicative factor $\frac{2}{\kappa}=1$.}
\be
P(z_3) = - \lim_{i\l_3 \rightarrow 0}i\l_3\bar\partial_3 G^+_{\D_3=1+i\l_3}(3)
\ee 
We would like to emphasize that \emph{this equation is true only in the OPE limit $z_3 \rightarrow z_4$}. \eqref{ssuper} is essentially the OPE \eqref{opesm} between the supertranslation current $P(z)$ and the graviton primary $G^+_{\D_4}(4)$,
\be
\begin{gathered}
P(z_3) G^+_{\D_4}(4) = \sum_{n=0}^{\infty} z_{34}^{n-1}  (P_{-n-1,-1}G^+_{\D_4})(4) \\ =
\frac{(P_{-1,-1}G^+_{\D_4})(4)}{z_3 - z_4} + (P_{-2,-1}G^+_{\D_4})(4) + (z_3 - z_4) (P_{-3,-1}G^+_{\D_4})(4) \\ + (z_3 - z_4)^2 (P_{-4,-1}G^{+}_{\D_4})(4) + ......
\end{gathered}
\ee

\section{Subleading Terms In The OPE}
The terms of $O(z_{34},\bar z_{34}, u_{34})$ in the expansion of the Mellin amplitude \eqref{OPEE} are given by,
\be
\begin{gathered}
\widetilde M_4(1^-2^-3^+4^+) = \langle G^-_{\D_1}(1)G^-_{\D_2}(2)G^{+}_{\D_3}(3)G^+_{\D_4}(4)\rangle \xrightarrow{z_3\rightarrow z_4} \ \supset  \\ - B(i\l_3,i\l_4)  \frac{\bar z_{34}}{z_{34}} \\ \times \Bigg\{ \frac{i\l_4 - i\l_3}{i\l_4 + i\l_3} \bigg(\frac{z_{34}}{z_{24}} + \frac{z_{34}}{z_{14}}\bigg) + \frac{i\l_3}{i\l_3 + i\l_4} \bigg( (1-i\l_1)\frac{z_{34}}{z_{24}} + (1-i\l_2)\frac{z_{34}}{z_{14}} + \frac{(2+ i\sum_{i=1}^4 \l_i)u_{12}z_{34}}{u_1z_{24} + u_4 z_{12} + u_2z_{41}}\bigg) \\ + \frac{i\l_3}{i\l_3+i\l_4} \bar z_{34} \frac{\partial}{\partial \bar z_4} \ - \frac{i\l_3}{i\l_3 + i\l_4} \frac{(2+ i\sum_{i=1}^4 \l_i) u_{34}z_{12}}{u_1z_{24} + u_4 z_{12} + u_2z_{41}} \Bigg\} \ i\frac{\partial}{\partial u_4} \widetilde M_3 (1^- 2^- 4^+)
   \end{gathered}
\ee
where the three point function $\widetilde M_3 (1^- 2^- 4^+) = \langle G^-_{\D_1}(1)G^-_{\D_2}(2)G^+_{\D_3 + \D_4 -1}(4)\rangle$ is defined in \eqref{3-point}. In the above expression we have obtained the $O(\bar z_{34})$ term by expanding the Dirac delta function appearing in the integrand in \eqref{OPEE}. We will now identify each term in the expansion as contribution coming from some descendant of the operator $G^{+}_{\D_3+\D_4-1}(4)$ : \\

1) The first term can be written as,
\be
\frac{i\l_4 - i\l_3}{i\l_4 + i\l_3} \bigg(\frac{z_{34}}{z_{24}} + \frac{z_{34}}{z_{14}}\bigg) i\frac{\partial}{\partial u_4} \widetilde M_3 (1^- 2^- 4^+) = \frac{i\l_4 - i\l_3}{i\l_4 + i\l_3} \ z_{34} \ \langle G^-_{\D_1}(1)G^-_{\D_2}(2)( P_{-2,-1}G^+_{\D_3 + \D_4 -1})(4)\rangle
\ee 

2) The second term can be written as,
\be\label{2nd}
\begin{gathered}
\frac{i\l_3}{i\l_3 + i\l_4} \bigg( (1-i\l_1)\frac{z_{34}}{z_{24}} + (1-i\l_2)\frac{z_{34}}{z_{14}} + \frac{(2+ i\sum_{i=1}^4 \l_i)u_{12}z_{34}}{u_1z_{24} + u_4 z_{12} + u_2z_{41}}\bigg) i\frac{\partial}{\partial u_4} \widetilde M_3 (1^- 2^- 4^+) \\ = \frac{i\l_3}{i\l_3 + i\l_4} \ z_{34} \frac{\partial}{\partial z_4} \  \langle G^-_{\D_1}(1)G^-_{\D_2}(2)(P_{-1,-1}G^+_{\D_3 + \D_4 -1})(4)\rangle \\ = \frac{i\l_3}{i\l_3 + i\l_4} \ z_{34} \  \langle G^-_{\D_1}(1)G^-_{\D_2}(2)( L_{-1}P_{-1,-1}G^+_{\D_3 + \D_4 -1})(4)\rangle
\end{gathered}
\ee

Now we would like to stress one important point. The expression for the three point function $\widetilde M_3(1^-2^-4^+)$ contains the kinematic theta function $\Theta(z_{42}/z_{12})\Theta(z_{14}/{z_{12}})$ and so, when $\partial/\partial z_{4}$ acts on it, it produces terms proportional to the delta functions $\delta(z_{42}/z_{12})$ and $\delta(z_{42}/z_{12})$. Now, as long as $z_{12}$ is finite, the delta function term is nonzero only if $z_4$ coincides with either $z_1$ or $z_2$. But this is ruled out in the OPE limit and so we should set the (contact) terms, obtained by differentiating the theta functions, to zero. We have taken this fact into account in coming from the first line to the second line in \eqref{2nd}. 

3) The third term can be written as,
\be
\frac{i\l_3}{i\l_3+i\l_4} \bar z_{34} \frac{\partial}{\partial \bar z_4} i\frac{\partial}{\partial u_4} \widetilde M_3 (1^- 2^- 4^+) = \frac{i\l_3}{i\l_3+i\l_4} \ \bar z_{34} \ \langle G^-_{\D_1}(1)G^-_{\D_2}(2)( \bar L_{-1}P_{-1,-1}G^+_{\D_3 + \D_4 -1})(4)\rangle
\ee

4) The fourth term can be written as, 
\be
\begin{gathered}
- \frac{i\l_3}{i\l_3 + i\l_4} \frac{(2+ i\sum_{i=1}^4 \l_i) u_{34}z_{12}}{u_1z_{24} + u_4 z_{12} + u_2z_{41}} \ i\frac{\partial}{\partial u_4} \widetilde M_3 (1^- 2^- 4^+) \\ = - i \frac{i\l_3}{i\l_3 + i\l_4} \ u_{34} \  \langle G^-_{\D_1}(1)G^-_{\D_2}(2)(P_{-1,-1}^2 G^+_{\D_3 + \D_4 -1})(4)\rangle
\end{gathered}
\ee \\

Therefore at the level of $4$-point function we get the following OPE,
\be\label{sope}
\begin{gathered}
G^+_{\D_3}(3)G^+_{\D_4}(4) \supset  - B(i\l_3,i\l_4)\frac{\bar z_{34}}{z_{34}} \Bigg( P_{-1,-1} + \frac{i\l_4 - i\l_3}{i\l_4 + i\l_3} z_{34} \ \boxed{P_{-2,-1}} + \frac{i\l_3}{i\l_3 + i\l_4} z_{34} \  L_{-1} P_{-1,-1} \\ + \bar z_{34} \frac{i\l_3}{i\l_3+i\l_4} \  \bar L_{-1}P_{-1,-1} - i u_{34} \frac{i\l_3}{i\l_3 + i\l_4} \ P_{-1,-1}^2 + \mathcal{O}(z^2_{3 4},\bar{z}^2_{34},z_{3 4} \bar{z}_{3 4})\Bigg) G^{+}_{\D_3 + \D_4 -1}(4)
\end{gathered}
\ee
We have boxed the descendant $P_{-2,-1}G^{+}_{\D_3 + \D_4 -1}(4)$ just to emphasize the fact that it is \emph{not} a Poincar\'e descendant. In the above equation we have not written an equality sign because some extra terms may be there which are not visible at the level of $4$-point function. In the following section we will see that this is not the case, according to BMS representation theory. 

\section{OPE Coefficients From BMS Algebra}
Let us start with the commutation relation between the supertranslation and (superrotation) Virasoro generators,
\be
[L_n, P_{a,b}] = \bigg(\frac{n-1}{2} - a\bigg) P_{a+n,b} \qquad   [\bar L_n, P_{a,b}] = \bigg(\frac{n-1}{2} - b\bigg) P_{a,b+n}
\ee
In particular for $n=0$ this reduces to,
\be
[L_0, P_{a,b}] = - \bigg(\frac{1}{2} + a\bigg) P_{a,b} \qquad   [\bar L_0, P_{a,b}] = - \bigg(\frac{1}{2} + b\bigg) P_{a,b}
\ee
We can see that the supertranslation generator $P_{a,b}$ has \emph{negative} holomorphic or antiholomorphic scaling dimension for $a>-1$ or $b>-1$ or both. So, if $\phi_{h,\bar h}$ is a primary operator then the dimension of the operator $[P_{a,b},\phi_{h,\bar h}(0)]$ can be made arbitrarily negative by choosing $a$ or $b$ to be a large positive integer. This motivates us to set

\be\label{hwt}
\boxed{
[P_{a,b} \ ,\phi_{h,\bar h}(0)] = 0 , \qquad a>-1 \ or \ b>-1 \ or \ both \ a,b > -1}
\ee   
for a \emph{BMS primary operator $\phi_{h,\bar h}$}. Now, for generators $P_{a,b}$ with $a > -1, b\ge -1$ or $a\ge -1, b>-1$, the condition \eqref{hwt} follows simply from the transformation law \eqref{sut1}. But there are generators $P_{-a,b}$ with $a>1, b>-1$ or $P_{a,-b}$ with $a>-1, b>1$, for which the condition \eqref{hwt} does not follow from the transformation law \eqref{sut1}. They generate singular supertranslation on the holomorphic side and non-singular supertranslation on the antiholomorphic side (and vice-versa). At this stage, except for the heuristic argument given below, the only justification for \eqref{hwt} for generators of this mixed type is that it gives the correct OPE coefficients, as we will see. 

The definition \eqref{hwt} of the BMS primary can be motivated by the following heuristic argument. Suppose the primary operator $\phi_{h,\bar h}(0)$ and (a subset of) its descendants\footnote{For $a>-1$ or $b>-1$ we should call the operators $[P_{a,b},\phi_{h,\bar h}(0)]$ "ascendants". But, for simplicity, we continue to call them descendants.} $[P_{a,b},\phi_{h,\bar h}(0)]$, appear in the OPE of two primaries $\phi_{h_1,\bar h_1}(u,z,\bar z)$ and $\phi_{h_2,\bar h_2}(0)$. The operator $[P_{a,b},\phi_{h,\bar h}(0)]$ appears in the OPE with a prefactor proportional to \\$z^{h-h_1-h_2 -(\frac{1}{2}+a)}\bar z^{\bar h-\bar h_1-\bar h_2 -(\frac{1}{2}+b)}$ \footnote{In general there will be powers of $u$ also, as can be seen from \eqref{OPEE} or \eqref{sope}, but we omit them for simplicity. It is obvious that the argument remains unchanged even if we include powers of $u$.}.\\ Now, symmetry requires \emph{all} descendants to be present in the OPE and this implies that the \emph{order of the pole} in $z$ or $\bar z$ \emph{cannot be bounded from above} because $a,b$ can be arbitrarily large positive integers. But, this is practically impossible because in the present situation the \emph{leading} pole can be obtained by Mellin transforming the splitting function in the collinear limit \cite{Pate:2019lpp,Fotopoulos:2019vac,Fan:2019emx}. So it seems reasonable to assume that the order of the pole in $z$ or $\bar z$ will be bounded from above and therefore, we should impose the condition \eqref{hwt} on the primary $\phi_{h,\bar h}$. 

For the Virasoro generators we have the standard conditions that,
\be\label{hwv1}
[L_n, \phi_{h,\bar h}(0)] = 0 , \qquad  [\bar L_n, \phi_{h,\bar h}(0)] = 0, \qquad n>0
\ee
and
\be\label{hwv1}
[L_0, \phi_{h,\bar h}(0)] = h \ \phi_{h,\bar h}(0) , \qquad  [\bar L_0, \phi_{h,\bar h}(0)] = \bar h \ \phi_{h,\bar h}(0)
\ee
Therefore we can now consider the descendants generated by the raising operators only. 

%So let us start with a primary $\phi_{h,\bar h}$. 

\subsection{A Hilbert space picture}
In the following discussion it will be convenient to assume a Hilbert space picture and define a vacuum $\ket{0}$ by the condition,
\be
P_{-1,-1}\ket{0} = 0
\ee
\be\label{mixedh}
P_{a,b}\ket{0} = 0 , \qquad a>-1 \ or \ b>-1 \ or \ both \ a,b > -1
\ee
\be
L_n\ket{0} = \bar L_n \ket{0} = 0 , \quad n\ge -1
\ee
We also define the state $\ket{h,\bar h}$ as,
\be
\ket{h,\bar h} = \phi_{h,\bar h}(0)\ket{0}
\ee
using the above definitions the BMS-primary state condition can be written as, 
\be\label{mixed}
P_{a,b}\ket{h,\bar h} = 0 , \qquad a>-1 \ or \ b>-1 \ or \ both \ a,b > -1
\ee
\be
L_n\ket{h,\bar h} = \bar L_n \ket{h,\bar h} = 0 , \quad n\ge 1
\ee
\be
L_0 \ket{h,\bar h} = h \ket{h,\bar h} , \quad \bar L_0 \ket{h,\bar h} = \bar h \ket{h,\bar h}
\ee 
For the restricted class of supertranslation generators $P_{a,b}$ with both $(a,b)>-1$, condition \eqref{mixedh} was also proposed in \cite{Bagchi:2016bcd}.
 
With this definition of the primary state the BMS descendants are given by,
\be
L_{-n_1}L_{-n_2}.......L_{-n_p} P_{-a_1,-b_1}P_{-a_2,-b_2}.......P_{-a_q,-b_q} \ket{h,\bar h}
\ee
where $n_1\ge n_2\ge.......n_p>0$ and $a_i >0, b_i>0$. We do not put any order on the $\{a_i\}$ and $\{b_i\}$ because the supertranslation generators commute among themselves. 

\subsection{Primary descendants}
Suppose $\ket{h,\bar h}$ is a BMS-primary state. Then it is easy to check using the commutation relations that the state $(P_{-1,-1})^n\ket{h,\bar h}$, with $n\ge 1$, is also a BMS-primary with weight $(h+n/2,\bar h+n/2)$. We denote this state by,
\be
(P_{-1,-1})^n \ket{h,\bar h} = \ket{h+n/2,\bar h+n/2}
\ee

\subsection{Graviton-Graviton OPE}
From now on our discussion will be confined to the OPE of two positive helicity gravitons denoted by $G^+_{\D}(u,z,\bar z)$. In our case, $\D=1+i\lambda$ and so,
\be
h = \frac{3+i\l}{2} , \quad \bar h = \frac{-1+i\l}{2}, \qquad \l\in \mathbb R
\ee
At this point let us note that although the Mellin amplitudes in Einstein gravity are UV divergent, the OPE \eqref{sope} has no singularity as $u_{34}\rightarrow 0$ and we can safely restrict the OPE to the celestial sphere at constant $u$, if we wish. From this point of view, the \emph{time coordinate $u$ acts as a covariant regulator in the celestial CFT}, which can be set to zero at the end of the calculation. For simplicity, we will do so. 

%Now, from \eqref{sope} we know that the leading term of the two graviton OPE is given by,
%\be
%G^+_{\D_1}(u,z,\bar z) G^+_{\D_2}(0) = - Bomorphic side and non-singular supertranslation on the antiholomorphic side (and vice-versa). At this stage, except for the heuristic argument given below, the only justification for \eqref{hwt} for generators of this mixed type is that it gives the correct OPE coefficients, as we will see. (i\l_1,i\l_2)\frac{\bar z}{z} \bigg( P_{-1,-1} + ............  \bigg) G^+_{\D_1 + \D_2 -1}(0)
%\ee
Once we set $u=0$, modulo the singular prefactor, the OPE expansion becomes a Taylor series in $z$ and $\bar z$ around the origin $(z=0,\bar z =0)$ and we can write the following \eqref{sope},
\be
G^+_{\D_1}(z,\bar z) G^+_{\D_2}(0) = -B(i\l_1,i\l_2)\frac{\bar z}{z} \bigg( P_{-1,-1} + \mathcal{O}(z,\bar{z}) \bigg) G^+_{\D_1 + \D_2 -1}(0)
\ee
where $G^+_{\D_1}(z,\bar z)=G^+_{\D_1}(u=0,z,\bar z)$. Let us now enumerate the possible subleading terms in the OPE at $O(z)$ and $O(\bar z)$. For this we remind ourselves that $P_{-1,-1}$ is a $(1/2,1/2)$ operator. Taking this into account we get, 

1) $O(z)$ operators : 
\be
L_{-1}P_{-1,-1} \rightarrow (3/2,1/2) , \quad P_{-2,-1} \rightarrow (3/2,1/2)
\ee 

2) $O(\bar z)$ operators : 
\be
\bar L_{-1}P_{-1,-1} \rightarrow (1/2,3/2) , \quad P_{-1,-2} \rightarrow (1/2,3/2)
\ee

If we keep powers of $u$ then there will be an additional operator at $O(u)$ given by $(P_{-1,-1})^2$, which follows from the fact that $u$ has scaling dimension $(-1/2,-1/2)$.

Now generically, at every order, there can be descendants and also \emph{new primaries}. But here we focus on the contribution arising from a specific primary $G^{+}_{\D_1 + \D_2 -1}(0)$. 

So let us write the first subleading order in the OPE as,
\be\label{opeh1}
\begin{gathered}
G^+_{\D_1}(z,\bar z) G^+_{\D_2}(0) = - B(i\l_1,i\l_2)\frac{\bar z}{z} \bigg( P_{-1,-1} + c_1 z  L_{-1}P_{-1,-1} + c_2 z P_{-2,-1} \\ + \bar c_1 \bar z \bar L_{-1} P_{-1,-1} + \bar c_2 \bar z P_{-1,-2}+ \mathcal{O}(z^2,\bar{z}^2,z \bar{z}) \bigg) G^+_{\D_1 + \D_2 -1}(0)
\end{gathered}
\ee
Here $\bar c_i$ is \emph{not} the complex conjugate of $c_i$. Also, for simplicity of notation, we have kept the dependence of the OPE coefficients $(c_i,\bar c_i)$ on the scaling dimensions of primary operators implicit. 

We will now compute the coefficients $(c_i,\bar c_i)$ from the fact that both sides of the OPE must transform in the same way under BMS transformation. In order to do this it is more convenient to write the OPE as, 
\be\label{opeh}
\begin{gathered}
G_{h_1,\bar h_1}(z,\bar z) \ket{h_2,\bar h_2} = - B(i\l_1,i\l_2)\frac{\bar z}{z} \bigg( P_{-1,-1} + c_1 z  L_{-1}P_{-1,-1} + c_2 z P_{-2,-1} \\ + \bar c_1 \bar z \bar L_{-1} P_{-1,-1} + \bar c_2 \bar z P_{-1,-2}+ \mathcal{O}(z^2,\bar{z}^2,z \bar{z}) \bigg) \ket{h_3,\bar h_3}
\end{gathered}
\ee
where,
\be\label{sd}
h_{1,2,3} = \frac{3+i\l_{1,2,3}}{2} , \quad \bar h_{1,2,3} = \frac{-1+i\l_{1,2,3}}{2} , \qquad \l_3 = \l_1 + \l_2
\ee
The BMS generators act on the operators as, 
\be
[L_n,\phi_{h,\bar h}(z,\bar z)] = \bigg[ z^{n+1} \partial + (n+1) h z^n \bigg] \phi_{h,\bar h}(z,\bar z)
\ee
\be
[\bar L_n,\phi_{h,\bar h}(z,\bar z)] = \bigg[ \bar z^{n+1} \bar\partial + (n+1) \bar h \bar z^n \bigg] \phi_{h,\bar h}(z,\bar z)
\ee
\be
[P_{a,b}, \phi_{h,\bar h}(z,\bar z)] = z^{a+1} \bar z^{b+1} \phi_{h+1/2, \bar h+1/2}(z,\bar z) = z^{a+1} \bar z^{b+1} (P_{-1,-1}\phi_{h, \bar h})(z,\bar z)
\ee
The first two relations are obtained by setting $u=0$ in commutation relations \eqref{cv}. The last commutator is obtained from,
\be
[P_{a,b}, \phi_{h,\bar h}(u,z,\bar z)] = z^{a+1}\bar z^{b+1} i\partial_u \phi_{h,\bar h}(u,z,\bar z)
\ee
by setting $u=0$ and recognizing that $i\partial_u\phi_{h,\bar h}(u,z,\bar z)$ is the BMS - primary (descendant) $(P_{-1,-1}\phi_{h,\bar h})(u,z,\bar z)$ with dimension $(h+1/2,\bar h+1/2)$. 

With this information one can readily compute the OPE coefficients, in the standard way, by applying the lowering operators to both sides of \eqref{opeh}. Let us now state the equations for the OPE coefficients obtained in this way, \\

1) Applying $P_{0,-1}$ we get,
\be\label{c1}
c_1 B(i\l_1,i\l_2) = B(i\l_1+1,i\l_2) \implies  c_1 = \frac{i\l_1}{i\l_1 + i\l_2}
\ee 

2) Applying $P_{-1,0}$ we get, 
\be\label{c1bar}
\bar c_1 B(i\l_1,i\l_2) = B(i\l_1+1,i\l_2) \implies  \bar c_1 = \frac{i\l_1}{i\l_1 + i\l_2}
\ee 

3) Applying $L_1$ we get, 
\be\label{c2}
(2h_3 + 1) c_1 + 2 c_2 = 2 h_1 -1 \implies c_2 = \frac{i\l_2 - i\l_1}{i\l_2 + i\l_1}
\ee

4) Applying $\bar L_1$ we get,
\be\label{c2bar}
(2\bar h_3 + 1) \bar c_1 + 2 \bar c_2 = 2 \bar h_1 +1 \implies \bar c_2 = 0
\ee 

This matches exactly with the OPE coefficients \eqref{sope}, obtained from graviton scattering amplitude, with the replacement $\l_3 \rightarrow \l_1$ and $\l_4 \rightarrow \l_2$. We would like to point out that since the descendants $P_{-2,-1}\ket{h_3,\bar h_3}$ and $P_{-1,-2}\ket{h_3,\bar h_3}$ appear in the OPE \eqref{opeh}, the above equations for the OPE coefficients probe the BMS algebra beyond the Poincar\'e subalgebra. 

Before we conclude, we would like to point out that to arrive at the equations for the OPE coefficients we need commutators of the form $[\bar L_{1}, P_{-2,-1}]\ket{h_3,\bar h_3} = P_{-2,0}\ket{h_3,\bar h_3}$, which by the definition \eqref{mixed} of a primary state is equal to zero. The generator $P_{-2,0}$ generates supertranslation of the mixed type, $u\rightarrow u + \epsilon \frac{\bar z}{z}$. So we can see that the condition that generators of the mixed type should \emph{also} annihilate a primary state is necessary to obtain the correct OPE coefficients starting from the BMS algebra, at least to the order we are working.  

\subsection{Virasoro Representations}
Since Virasoro algebra is a subalgebra of the BMS algebra, one should be able to decompose a BMS representation into irreducible Virasoro representations. This is useful because in this way one can use the known results from the (highest-weight) Virasoro representation theory to compute some of the OPE coefficients. At the first subleading order this can be done in the following way. 

Let us write the OPE at the first subleading order as,
\be\label{opes2}
\begin{gathered}
G_{h_1,\bar h_1}(z,\bar z) \ket{h_2,\bar h_2} = - B(i\l_1,i\l_2)\frac{\bar z}{z} \bigg( P_{-1,-1} + c_1 z  L_{-1}P_{-1,-1} + c_2 z P_{-2,-1} \\ + \bar c_1 \bar z \bar L_{-1} P_{-1,-1} + \mathcal{O}(z^2,\bar{z}^2, z \bar{z})  \bigg) \ket{h_3,\bar h_3}
\end{gathered}
\ee
where we have set $\bar c_2=0$ following \eqref{c2bar}. 

Among these states, the state $\bar L_{-1}P_{-1,-1}\ket{h_3,\bar h_3}$ is the antiholomorphic Virasoro descendant of the BMS primary $P_{-1,-1}\ket{h_3,\bar h_3}$. We have two more states, given by $L_{-1}P_{-1,-1}\ket{h_3,\bar h_3}$ and $P_{-2,-1}\ket{h_3,\bar h_3}$, one of which is a Virasoro descendant and the other one is a supertranslation descendant. So let us consider the state,
\be
\boxed{
\ket{\psi} = \bigg( P_{-2,-1} - \frac{2}{2h_3 +1} L_{-1} P_{-1,-1} \bigg)\ket{h_3,\bar h_3}}
\ee
The corresponding field can be written as,
\be
\psi(z,\bar z) = (P_{-2,-1}G_{h_3,\bar h_3})(z,\bar z) - \frac{2}{2h_3 +1}(L_{-1} P_{-1,-1} G_{h_3,\bar h_3})(z,\bar z)
\ee
where the correlation function with the insertion of $(P_{-2,-1}G_{h_3,\bar h_3})(z,\bar z)$ is given by \eqref{nou}. Now one can check using the BMS algebra and the definition of the BMS primary state $\ket{h_3,\bar h_3}$ that $\ket{\psi}$ is a \emph{Virasoro primary} but, \emph{not} a BMS primary. For example, one can check that 
\be
P_{0,-1}\ket{\psi}\propto P_{-1,-1}^2 \ket{h_3,\bar h_3} \ne 0
\ee
and so on. Now using the \emph{Virasoro primary} $\ket{\psi}$ we can rewrite the OPE \eqref{opes2} as, 
\be\label{vdec}
\begin{gathered}
G_{h_1,\bar h_1}(z,\bar z) \ket{h_2,\bar h_2} = - B(i\l_1,i\l_2)\frac{\bar z}{z} \Bigg( P_{-1,-1} + c_1' z  L_{-1}P_{-1,-1} + \bar c_1 \bar z \bar L_{-1} P_{-1,-1} + ... \Bigg) \ket{h_3,\bar h_3} \\- B(i\l_1,i\l_2) \bar z c_2 \ket{\psi} + ...... \\ = - B(i\l_1,i\l_2)\frac{\bar z} {z} \bigg( 1 + c_1' z  L_{-1} + \bar c_1 \bar z \bar L_{-1} + ... \bigg) \ket{h_3+1/2,\bar h_3+1/2} \\- B(i\l_1,i\l_2) c_2 \ \bar z \ket{\psi} + ......
\end{gathered}
\ee
where,
\be\label{c1'}
c_1' = c_1 + \frac{2}{2h_3 +1} c_2
\ee
The OPE \eqref{vdec} now has the familiar structure of OPE in $2$-D CFT. The first line consists of the Virasoro primary $\ket{h_3+1/2,\bar h_3+1/2}$ and its level $1$ descendants. In the second line we have a new Virasoro primary $\ket{\psi}$ with the (Virasoro) structure constant $B(i\l_1,i\l_2) c_2$. The Virasoro structure constant $B(i\l_2,i\l_2)c_2$ cannot be determined by Virasoro algebra alone but, the coefficients $c_1'$ and $\bar c_1$ are determined by Virasoro algebra. They are given by the known formulas,

\be\label{v1}
c_1' = \frac{h_3' + h_1 - h_2}{2h_3'}, \qquad h_3' = h_3 + \frac{1}{2}
\ee
\be\label{v2}
\bar c_1 = \frac{\bar h_3' + \bar h_1 - \bar h_2}{2\bar h_3'}, \qquad \bar h_3' = \bar h_3 + \frac{1}{2}
\ee
Using the values \eqref{sd} for the scaling dimensions we get, from \eqref{v1} and $\eqref{v2}$, that
\be
c_1' = \frac{2+i\l_1}{4+i(\l_1+\l_2)}
\ee
and
\be
\bar c_1 = \frac{i\l_1}{i\l_1 + i\l_2}
\ee
We can see that the value of $\bar c_1$ obtained in this way using Virasoro algebra matches with the value \eqref{c1bar} obtained using translational invariance of the OPE. Similarly, the value of $c_1'$ also matches with \eqref{c1'} once we substitute the values \eqref{c1} and \eqref{c2} of $c_1$ and $c_2$, obtained using BMS algebra. This matching is a check of the overall consistency of the procedure.

\section{Future Directions}
The problem of extracting the celestial OPE from flat space scattering amplitudes consists of two parts. The first part is the (holomorphic or antiholomorphic) collinear expansion including subleading terms and the second part is the determination of the celestial correlation functions with the insertion of BMS descendants. The correlation functions involving superrotation or Virasoro descendants are well known from the work of Belavin-Polyakov-Zamolodchikov on two dimensional CFTs. The new objects are the supertranslation descendants. Among these, the simplest ones are the descendants $\{P_{-a,-1}\phi\ , a>1\}$ created by singular (anti) holomorphic supertranslations. These are captured in a straightforward manner by supertranslation Ward-identity \eqref{supward1} following from Weinberg's soft graviton theorem. The correlation functions with the insertion of (anti) holomorphic supertranslation descendants follow from this Ward-identity and is given by \eqref{superd} or \eqref{nou}. But, there are also descendants of the form $\{P_{-a,-b}\phi , a>1, b>1\}$ created by supertranslations, which are neither purely holomorphic nor antiholomorphic. For example, if we want to go to higher order in the OPE expansion \eqref{opeh1}, then at $O(z\bar z)$ we encounter the descendant $P_{-2,-2}G^+_{\D_1+\D_2-1}(0)$. Unless we know the Mellin amplitude with the insertion of this descendant, it will not be possible to extract the OPE at higher order. Although the Ward-identity \eqref{supward1} captures all the supertranslation descendants \cite{Strominger:2013jfa} - not just holomorphic - it may require a more involved procedure to determine the correlation function of a general supertranslation descendant. 

Another important point is that the BMS algebra may have (field-dependent) central extension \cite{Barnich:2017ubf,Distler:2018rwu}. In this paper the central extension does not play any role because in the first subleading order the Virasoro descendants $\{ L_{-n}\phi , n>1\}$ do not appear, although they appear in the higher order of the OPE. The values of the OPE coefficients should depend on the central charge and perhaps one can determine the central charge by demanding that the OPE coefficients determined from the BMS algebra match with those obtained from the subleading terms in the collinear expansion of the Mellin amplitude. We hope to return to these problems in future.   

\section{Acknowledgements}
SB would like to thank the participants of "Saha Theory Workshop 2020: Amplitudes and Correlators" where a part of this work was presented. SB would also like to thank Arjun Bagchi, Justin David, Pranjal Pandey, Partha Paul and Arnab Priya Saha for helpful discussion. SG would like to thank Yasha Neiman and Vyacheslav Lysov for helpful discussions. SG is supported by the Quantum Gravity Unit of the Okinawa Institute of Science and Technology (OIST). RG would like to thank Guy Jehu, Marius de Leeuw and Tristan McLoughlin for useful discussions. The work of RG has received funding from the European Union's Horizon 2020 research and innovation programme under the Marie Sklodowska-Curie grant agreement No. 764850. The work of SB is partially supported by the Science and Engineering Research Board (SERB) grant MTR/2019/000937 (Soft-Theorems, S-matrix and Flat-Space Holography).

\section{Appendix}

\subsection{Mellin Transform of Four Graviton Amplitude and Holomorphic Collinear Expansion}
In this appendix we discuss the Mellin transform of the tree-level four graviton scattering amplitude in Einstein theory in detail. The four graviton amplitude in momentum space is given by, 
\be
\begin{gathered}
M_4(1^-2^-3^+4^+) \\ = \frac{\langle12\rangle^7[12]}{\langle13\rangle \langle14\rangle\langle23\rangle\langle24\rangle\langle34\rangle^2} \ \delta^4 (\omega_1q(z_1,\bar z_1)+\omega_2q(z_2,\bar z_2)-\omega_3q(z_3,\bar z_3)-\omega_4q(z_4,\bar z_4)) \\
= - 4 \frac{\om_1^2\om_2^2}{\om_3\om_4}\frac{z_{12}^6 \bar z_{34}}{z_{13}z_{14}z_{23}z_{24}z_{34}} \ \delta^4 (\omega_1q(z_1,\bar z_1)+\omega_2q(z_2,\bar z_2)-\omega_3q(z_3,\bar z_3)-\omega_4q(z_4,\bar z_4))
\end{gathered}
\ee
where we have taken $(1,2)$ to be incoming and $(3,4)$ to be outgoing. We work in split signature so that we can treat $z$ and $\bar z$ as independent real variables. The OPE limit, we are interested in, is $z_3 \rightarrow z_4$ with $\bar z_{34}$ held fixed. 

The corresponding Mellin amplitude is given by, 
\be
\begin{gathered}
\widetilde M_4(1^-2^-3^+4^+) \\ = \int_{0}^{\infty}d\omega_1\int_{0}^{\infty}d\omega_2\int_{0}^{\infty}d\omega_3\int_{0}^{\infty}d\omega_4 \ \omega_1^{i\l_1} \omega_2^{i\l_2}\omega_3^{i\l_3}\omega_4^{i\l_4} e^{-i\sum_{i=1}^4 \epsilon_i \omega_i u_i} M_4(1^-2^-3^+4^+)
\end{gathered}
\ee
where $\epsilon_i=\pm 1$ for outgoing and incoming particles, respectively. Here we take $(1,2)$ to be incoming and $(3,4)$ to be outgoing and so, $\epsilon_1=\epsilon_2 =-1$ and $\epsilon_3 = \epsilon_4 = 1$.

Now, using the momentum conserving delta function we solve for $\omega_1$, $\omega_2$, $\bar z_{14}$ and $\bar z_{24}$ and write,
\be
\begin{gathered}
\delta^4 (\omega_1q(z_1,\bar z_1)+\omega_2q(z_2,\bar z_2)-\omega_3q(z_3,\bar z_3)-\omega_4q(z_4,\bar z_4)) \\ = \frac{1}{4\om_1^*\om_2^*} \frac{1}{z_{12}^2} \delta (\om_1 - \om_1^*) \delta(\om_2 - \om_2^*) \ \delta\left(\bar z_{14} + \frac{\om_3}{\om_1^*} \frac{z_{23}}{z_{12}} \bar z_{34}\right) \delta \left(\bar z_{24} - \frac{\om_3}{\om_2^*} \frac{z_{13}}{z_{12}} \bar z_{34}\right) 
\end{gathered}
\ee
where
\be
\om_1^* = - \frac{z_{24}}{z_{12}} (\om_3 + \om_4) + \frac{z_{34}}{z_{12}}\om_3
\ee
\be
\om_2^* =  \frac{z_{14}}{z_{12}} (\om_3 + \om_4) - \frac{z_{34}}{z_{12}}\om_3
\ee
Now we make a change of variable 
\be
\om = \om_3 + \om_4, \quad \om_3 = t\om , \quad \om_4 = (1-t)\om, \qquad 0\le t \le1
\ee
In terms of $\omega$ and $t$,
\be
\om_1^* =\om\bigg(- \frac{z_{24}}{z_{12}} + t \frac{z_{34}}{z_{12}}\bigg)
\ee
\be
\om_2^* = \om\bigg( \frac{z_{14}}{z_{12}} - t \frac{z_{34}}{z_{12}}\bigg)
\ee
Now since $\om_1$ and $\om_2$ are positive, $\om_1^*$ and $\om_2^*$ must also be positive. In the OPE limit this is indeed true. To see this, let us first note that $\om_1^*$ and $\om_2^*$ are zero when,
\be
\om_1^* = 0 \implies t_1^* = \frac{z_{24}}{z_{34}}
\ee
and 
\be
\om_2^* = 0 \implies t_2^* = \frac{z_{14}}{z_{34}}
\ee
In the OPE limit, $|z_{34}| \ll |z_{13}|,|z_{23}|,|z_{14}|,|z_{24}|$ and so $|t_1^*|,|t_2^*| \gg1$, which is outside the range of $t$. Therefore $\om_1^*$ and $\om_2^*$ do not change sign as $t$ runs from $0$ to $1$.  

Now the integrals over $\om_1$ and $\om_2$ gives rise to the following theta functions
\be
\int_{0}^{\infty} \om d\om \int_{0}^{1} dt \ \Theta(\om_1^*)\Theta(\om_2^*) ..............
\ee
Since $\om_1^*$ and $\om_2^*$ do not change sign in the interval $0\le t \le1$, the theta functions are constant and we take them outside the integral and evaluate at $t=0$. This gives us 
\be\label{theta}
\Theta\bigg(\frac{z_{42}}{z_{12}}\bigg) \Theta\bigg(\frac{z_{14}}{z_{12}}\bigg) \int_{0}^{\infty} \om d\om \int_{0}^{1} dt ...........
\ee
where we have also used the fact that $\om>0$. Now, the theta functions appearing in \eqref{theta} are precisely the ones appearing in the three point function $\langle1^-2^-4^+\rangle$ where $1,2$ are incoming and $4$ is outgoing. The rest of the integrals is straightforward and give rise to the Mellin transform \eqref{OPEE} in the OPE limit. 

\subsection{Virasoro Ward Identity in Modified Mellin Basis}

This section of the appendix is based on results from \cite{Partha}. 
The existence of a Virasoro symmetry for 3+1 quantum gravity in asymptotically flat spacetime is based on the subleading soft graviton theorem. This was first shown in \cite{Kapec:2014opa, Kapec:2016jld} for four dimensions at semiclassical level and later generalized to loop level \cite{He:2017fsb}. We want here to establish analogous results in the modified Mellin basis, which will confirm the structure of the conformal transformations we assumed in the paper. The modified Mellin transform of the creation (resp. annihilation) operator $a(p(\omega,z,\bar{z}),\sigma)$ (resp. $a^{\dagger}(p(\omega,z,\bar{z}),\sigma)$) is given by \cite{Banerjee:2018gce},
\begin{align}
A(u,z,\bar{z},\lambda,\sigma) &= N \int_0^{\infty} d \omega \, \omega^{-i \lambda} e^{-i \omega u} a(p(\omega,z,\bar{z}),\sigma) \\
A^{\dagger}(u,z,\bar{z},\lambda,\sigma) &= N \int_0^{\infty} d \omega \, \omega^{i \lambda} e^{i \omega u} a^{\dagger}(p(\omega,z,\bar{z}),\sigma)
\label{eqn:Mellin_a}
\end{align}
where $N$ is a normalization constant which we will take to be $\frac{1}{\sqrt{8 \pi^2}}$. If we do a formal Laurent expansion of $a(p(\omega,z,\bar{z}),\sigma)$ in $\omega$ we will have
\begin{align}
a(\omega,z,\bar{z},\sigma) = \sum_n \frac{\mathcal{S}_{1-n}(z,\bar{z},\sigma)}{\omega^n}
\label{eqn:soft_exp}
\end{align}
One can easily check using the Lorentz transformation property of $a(\omega,z,\bar{z},\sigma)$ that the coefficients are "soft operators" which transform as primary of weight $n$ and spin $\sigma$ under the conformal group $SL(2,\mathbb{C})$. Let us now substitute the expansion \eqref{eqn:soft_exp} into \eqref{eqn:Mellin_a}. Doing the integral we get
\begin{align}
A(u,z,\bar{z},\lambda,\sigma) = N \sum_n \frac{\Gamma(2 - \Delta - n)}{(i u)^{2 - \Delta - n}} \mathcal{S}_{1-n}(z,\bar{z},\sigma)
\end{align}
where $u = u - i \delta$ and $\delta \to 0^+$. Let us notice that this integral is ill-defined for any $n \neq 1$ if the exponential factor $e^{-i \omega u}$ is not there.
For the time being let us assume that $n \leq 1$. Now as $\lambda \to 0$ the singularity comes only from the $n=1$ term and it is a pole of the $\Gamma$ function. Around the pole we can write
\begin{align}
A(u,z,\bar{z},\lambda,\sigma) = -N \frac{1}{i \lambda} \mathcal{S}_0(z,\bar{z},\sigma) + \text{finite terms}
\end{align}
Therefore the residue at the pole is the leading soft operator $\mathcal{S}_0(z,\bar{z},\sigma)$, up to a sign. So we can write
\begin{align}
\mathcal{S}_0(z,\bar{z},\sigma) = -\lim_{\Delta \to 1} (\Delta - 1) \left[ (i u)^{\Delta - 1} \frac{1}{N} A(u,z,\bar{z},\lambda, \sigma) \right]
\label{eqn:S1_subleading}
\end{align}
When inserted in an S-matrix element this leads to the leading conformal soft theorem. We would like to stress that this construction is valid only when all the operators are inserted in an S-matrix element and then we can meaningfully talk about analytic continuation in $\lambda$. In the same fashion we can talk about subleading ($n = 0$) conformal soft limit which is obtained in the limit $\lambda \to i$: we have
\begin{align}
\mathcal{S}_{1}(z,\bar{z},\sigma) = \lim_{\Delta \to 0} \Delta \left(1 - u \frac{\partial}{\partial u} \right) \left[ (i u)^{\Delta} \frac{1}{N} A(u,z,\bar{z},\lambda, \sigma) \right]
\end{align}
The insertion of such an operator will lead to the subleading conformal soft theorem without the contamination from the leading soft theorem. It is worth noticing that the appearance of a single power of $u$ is determined by dimensional analysis because $u$ transforms like a primary of dimension $-1$ and spin $0$. Similarly we can write for $p \geq 1$
\begin{align}
\mathcal{S}_{p+1}(z,\bar{z},\sigma) = \lim_{\Delta \to -p} (-1)^{p} \frac{\Delta + p}{(1 + p)!} \prod_{n = 0}^p \left(1 - n - u \frac{\partial}{\partial u} \right) \left[(i u)^{\Delta + p} \frac{1}{N} A(u,z,\bar{z},\lambda, \sigma) \right]
\end{align}
We can see that in the absence of $n > 1$ terms in the soft expansion \eqref{eqn:soft_exp} the poles in the upper-half $\lambda$ plane correspond to the IR behaviour of the scattering amplitude. 

We're ready now to show how the Virasoro Ward identity arises from the subleading soft theorem. Let $\mathcal{A}_{n+1}(\{p_i\}_{i \in \{1,...,n\}},q)$ be the $n+1$ scattering amplitude involving $n$ massless particles and one graviton of momentum $q^{\mu}$ and polarization $\epsilon_{\mu \nu}^{(\pm)}(q)$. The soft limit for the graviton of momentum $q \to 0$ gives \cite{Weinberg:1965nx,Cachazo:2014fwa}
\begin{align}
\mathcal{A}_{n+1}(\{p_i\}_{i \in \{1,...,n\}},q) \to \left[S_0^{(\pm)} + S_1^{(\pm)} + \mathcal{O}(q)\right] \mathcal{A}_{n}(\{p_i\}_{i \in \{1,...,n\}})
\end{align}
where $\mathcal{A}_{n}(\{p_i\}_{i \in \{1,...,n\}})$ is the amplitude without the soft graviton and\footnote{ We are using here the convention $\kappa = \sqrt{32 \pi G} = 2$ as done in the main text of the paper.}
\begin{align}
S_0^{(\pm)} &=  \sum_{j=1}^n \epsilon_j \frac{p_j^{\mu} p_j^{\nu} \epsilon_{\mu \nu}^{(\pm)}(q)}{p_j \cdot q} \\
S_1^{(\pm)} &=  -i \sum_{j=1}^n \epsilon_j \frac{ \epsilon_{\mu \nu}^{(\pm)}(q) p_j^{\mu}}{p_j \cdot q} q_{\lambda} J^{\lambda \nu,(\pm)}_j 
\end{align}
where $J^{\lambda \nu,(\pm)}_j$ is the sum of the spin and angular momentum of the $j$-th particle. Let's focus, without loss of generality, on the minus helicity case. In the coordinates $(\omega,z,\bar{z})$ the subleading soft factor becomes
\begin{align}
S_1^{(-)} =  \sum_{j=1}^n \epsilon_j \left[\frac{(z - z_j)^2}{(\bar{z} - \bar{z}_j)} \left( \frac{2 \hat{h}_j}{(z - z_j)} - \partial_{z_j} \right) \right]
\end{align}
where $\hat{h}_j = \frac{1}{2} (\sigma_j -\omega_j \partial_{\omega_j})$. As we have already explained the insertion of the operator $\mathcal{S}_1$ defined in \eqref{eqn:S1_subleading} in the scattering amplitudes will extract the subleading soft behaviour. Now supposing that we have $n_1$ incoming  and $n_2$ outgoing hard particles ($n_1 + n_2 = n$) we can define the subleading soft graviton contribution to a $(n+1)$-point scattering amplitude in the modified Mellin basis as follows
\begin{align}
%& \mathcal{A}_{n+1}(\{p_i\}_{i \in \{1,...,n\}},q)\Big|_{\mathcal{O}(q^0)} = \nonumber \\
&  \bra{0} \mathcal{S}_1(z,\bar{z},-) \left[ \prod_{j_2=1}^{n_2} \, A_{\text{out}}(u_{j_2},z_{j_2},\bar{z}_{j_2},\lambda_{j_2},\sigma_{j_2}) \right] \left[ \prod_{j_1=1}^{n_1} \, A^{\dagger}_{\text{in}}(u_{j_1},z_{j_1},\bar{z}_{j_1},\lambda_{j_1},\sigma_{j_1}) \right] \ket{0} 
\end{align}
so that
\begin{align}
&  \lim_{\Delta \to 0} \Delta \left(1 - u \frac{\partial}{\partial u} \right) (i u)^{\Delta} \left[\prod_{j_2=1}^{n_2} \int_0^{\infty} d \omega_{j_2} \, (\omega_{j_2})^{- i \lambda_{j_2}} e^{-i \omega_{j_2} u_{j_2}} \right] \left[\prod_{j_1=1}^{n_1} \int_0^{\infty} d \omega_{j_1} \, (\omega_{j_1})^{i \lambda_{j_1}} e^{i \omega_{j_1} u_{j_1}} \right] \times \nonumber \\
& \times \frac{1}{N} \bra{0} A_{\text{out}}(u,z,\bar{z},\lambda, -) \, \prod_{j_2=1}^{n_2}  a_{\text{out}}(p_{j_2}(u_{j_2},z_{j_2},\bar{z}_{j_2}),\sigma_{j_2}) \, \prod_{j_1=1}^{n_1}  a_{\text{in}}^{\dagger}(p_{j_1}(u_{j_1},z_{j_1},\bar{z}_{j_1}),\sigma_{j_1}) \ket{0} = \nonumber \\
&= \left[\prod_{j_2=1}^{n_2} \int_0^{+\infty} d \omega_{j_2} \, (\omega_{j_2})^{- i \lambda_{j_2}} e^{-i \omega_{j_2} u_{j_2}} \right] \left[\prod_{j_1=1}^{n_1} \int_0^{+\infty} d \omega_{j_1} \, (\omega_{j_1})^{i \lambda_{j_1}} e^{i \omega_{j_1} u_{j_1}} \right] S_1^{(-)} \mathcal{A}_{n}(\{p_i\})
\end{align}
as expected. At this point it is worth considering the following operator \cite{Kapec:2016jld}
\begin{align}
T(z) = \frac{1}{2\pi} \int d^2 w \frac{1}{z - w} \partial^3_{w} \mathcal{S}_1(w,\bar{w},-)
\end{align}
If we identify the correlators of the dual theory with S-matrix elements transformed to the modified Mellin basis as follows
\begin{align}
\langle \prod_{i = 1}^n \phi_{\Delta_i, \sigma_i}(z_i, \bar{z}_i, u_i) \rangle = \prod_{i = 1}^n\int_0^{\infty} d \omega_i \, \omega_{i}^{\Delta_i - 1} e^{-i \epsilon_i \omega_i u_i} \mathcal{A}(p(\omega_i,z_i, \bar{z}_i),\sigma) 
\end{align}
then the insertion of such an operator $T(z)$ will give, after some algebra,\footnote{we have used the the following identity $\partial_{\bar{w}} \frac{1}{z - w} = 2 \pi \delta^2(z - w)$}
\begin{align}
&\langle T(z) \phi_{h_1 \bar{h}_1}(u_1, z_1,\bar{z}_1) ... \phi_{h_n \bar{h}_n}(u_n, z_n,\bar{z}_n) \rangle = \nonumber \\ 
&= \sum_{i=1}^n \left[ \frac{h_i + \frac{1}{2} u_i \partial_{u_i}}{(z - z_i)^2} + \frac{1}{z - z_i}  \partial_{z_i} \right] \langle \phi_{h_1 \bar{h}_1}(u_1, z_1,\bar{z}_1) ... \phi_{h_n \bar{h}_n}(u_n, z_n,\bar{z}_n) \rangle
\end{align}
where $h_i = \frac{\Delta_i + \sigma_i}{2}$. A similar calculation can be done for the positive helicity graviton using the following operator
\begin{align}
\overline{T}(\bar{z}) = \frac{1}{2 \pi} \int d^2 w \frac{1}{\bar{z} - \bar{w}} \partial^3_{\bar{w}} \mathcal{S}_1(w,\bar{w},+)
\end{align}
and after similar steps we end up with 
\begin{align}
&\langle \overline{T}(\bar{z}) \phi_{h_1 \bar{h}_1}(u_1, z_1,\bar{z}_1) ... \phi_{h_n \bar{h}_n}(u_n, z_n,\bar{z}_n) \rangle = \nonumber \\ 
&= \sum_{i=1}^n \left[ \frac{\bar{h}_i + \frac{1}{2} u_i \partial_{u_i}}{(\bar{z} - \bar{z}_i)^2} + \frac{1}{\bar{z} - \bar{z}_i}  \partial_{\bar{z}_i} \right] \langle \phi_{h_1 \bar{h}_1}(u_1, z_1,\bar{z}_1) ... \phi_{h_n \bar{h}_n}(u_n, z_n,\bar{z}_n) \rangle
\end{align}
where $\bar{h}_i = \frac{\Delta_i - \sigma_i}{2}$.

%\iffalse

\end{document}